\documentclass[twocolumn,pre,aps,showpacs,amsmath]{revtex4}
\usepackage{graphicx}

\begin{document}

\title{Heterogeneity-Induced Inhibitory Coherence in An Ensemble of Subthreshold and Suprathreshold Type-I Neurons}
\author{Duk-Geun Hong and Sang-Yoon Kim}
\thanks{Corresponding Author}
\email{sykim@kangwon.ac.kr}
\affiliation{Department of Physics, Kangwon National University, Chunchon, Kangwon-Do 200-701, Korea}
\author{Woochang Lim}
\affiliation{Department of Science Education, Daegu National University of Education, Daegu 705-115, Korea}

\begin{abstract}
We study inhibitory coherence (i.e., collective coherence by synaptic inhibition) in an ensemble of globally-coupled type-I neurons which can fire at arbitrarily low frequencies. No inhibitory coherence is observed in a homogeneous ensemble composed of only subthreshold neurons (which cannot fire spontaneously without noise). By increasing the fraction of (spontaneously firing) suprathreshold neurons $P_{supra}$, heterogeneity-induced inhibitory
coherence is investigated in a heterogeneous ensemble of subthreshold and suprathreshold neurons. As $P_{supra}$ passes a threshold $P_{supra}^*$, suprathreshold neurons begin to synchronize and play the role of coherent inhibitors for the emergence of inhibitory coherence. Thus, regularly-oscillating ensemble-averaged global potential appears for $P_{supra} > P_{supra}^*$. For this coherent case suprathreshold neurons exhibit coherent mixed-mode oscillations with a fast subthreshold (small-amplitude) hopping frequency and a lower spiking frequency. By virtue of their coherent inhibition, sparsely synchronized suprathreshold neurons suppress noisy activities of subthreshold neurons. Thus, only coherent subthreshold hoppings appear in the individual potentials of subthreshold neurons. We also characterize the inhibitory coherence in terms of the ``statistical-mechanical'' spike-based and correlation-based measures and find that the degree of inhibitory coherence increases with increasing $P_{supra}$ for $P_{supra} > P_{supra}^*$. Finally, effect of sparse randomness of synaptic connectivity on the inhibitory coherence and universality of the
heterogeneity-induced inhibitory coherence are briefly discussed.
\end{abstract}


\maketitle

\section{Introduction}
\label{sec:INT}

Recently, much attention has been paid to rhythms of the brain \cite{Buz}. Coherence of neural oscillations may be used for efficient sensory and cognitive processing (e.g., feature integration, selective attention, working memory, and decision making) \cite{W_Review,Gray}. This kind of neural coherence is also correlated with pathological rhythms associated with neural diseases (e.g., epileptic seizures and tremors in the Parkinson's disease) \cite{ND}. Here, we are interested in these coherent brain rhythms. A brain circuit is composed of a few types of excitatory principal cells and diverse types of inhibitory interneurons. Interneuron diversity increases the computational power of principal cells \cite{Buz}. Effect of chemical synapses on coherent brain rhythms has been much investigated in neural systems composed of excitatory and/or inhibitory neurons \cite{W_Review,Wang}. Historically, recurrent excitation between principal cells is the conventional coherence mechanism \cite{Ex}. However, when the decay time of the synaptic interaction is enough long, mutual inhibition between interneurons (rather than excitation) may synchronize individual neural firings \cite{Stab1,Stab2}. By providing a coherent oscillatory output to the principal cells, interneuronal networks play the role of the backbones (i.e., pacemakers) of many brain rhythms such as the thalamocortical spindle rhythms \cite{WR,GR} and the fast gamma rhythms in the hippocampus and the neocortex \cite{WB,White,ING,Gamma1}. When the feedback between the excitatory and the inhibitory populations is strong, neural coherence occurs via the ``cross-talk'' between the two populations \cite{ING,Gamma1,HM,BK}. In these computational studies of neural coherence, different types of network architectures have been considered \cite{W_Review}; all-to-all networks where every neuron is coupled to every other neuron, sparse random networks where synaptic connections are sparse, and complex networks such as small-world networks (with predominantly local connections and rare long-distance connections) \cite{SW} and scale-free networks (with a few percent of hub neurons with an exceptionally large number of connections) \cite{SFN}.

Neurons in the neural system exhibit a variety of morphological and physiological properties. However, close to threshold, this remarkable richness may be grouped broadly into two basic types of excitability, often referred to as type I and type II \cite{Ex1}. When the strength of a constant input current passes a threshold, type-I neurons can fire at arbitrarily low frequencies and they can smoothly encode the strength of the input into the output firing frequency. In contrast, type-II neurons have a non-zero minimum frequency of firing and they fire in a narrow frequency band which is relatively insensitive to changes in the strength of the applied current. Different types of excitability occur because neurons have different bifurcations of resting and spiking states \cite{Ex2}. For the type I neurons, oscillations emerge via a
saddle-node bifurcation on an invariant circle. As the bifurcation parameter (i.e., strength of the injected current) passes a threshold, the stable and the unstable fixed points coalesce and then disappear, leaving a large-amplitude stable periodic orbit. This is a global bifurcation and the frequency of the global loop can be arbitrarily small. On the other hand, for type-II neurons a transition from a resting state to a periodically spiking state occurs through a Hopf bifurcations with a finite non-zero firing frequency. According to their bifurcations, neurons may also be classified into integrators and resonators \cite{Ex3}. Type-I neurons act as integrators without subthreshold oscillations, and they prefer high-frequency input: the higher the frequency of the input, the sooner they fire. In contrast, type-II neurons exhibit damped subthreshold oscillations and act as resonators: they prefer oscillatory input with the same frequency as that of damped oscillations. According to their excitability type, neurons make distinctly different responses to stimuli which have important implications for their distinct roles in generating population rhythms \cite{Hansel,Type1,Izhi,Type2,India}.

In this paper, we study inhibitory coherence (i.e., collective coherence by synaptic inhibition) in an ensemble of globally-coupled type-I neurons. Neural models exhibiting the type-1 excitability include the Connor model for the crab leg axons \cite{Connor}, the Wang-Buzsaki model for inhibitory interneurons \cite{WB}, the Hindmarsh-Rose model \cite{HR}, and the Morris-Lecar (ML) model \cite{ML1} under some circumstances. In Section \ref{sec:HE}, we describe the biological conductance-based ML neuron model with voltage-gated ion channels.
The ML neurons (used in our study) exhibit the type-I excitability, and they
interact via inhibitory GABAergic synapses whose activity increases fast and decays slowly. Inhibitory coherence (which is our main concern) is particularly important because it plays a significant role in integration of sensory and cognitive information; for example, impaired inhibitory coherence is believed to be associated with schizophrenia and attention deficit disorder \cite{IC1,IC2}. Hence, it is important to understand mechanisms for the emergence of inhibitory coherence. Many works exploring mechanisms of neural coherence were done in neural systems composed of spontaneously firing (i.e., self-oscillating) suprathreshold neurons (above the threshold) \cite{W_Review,Wang}. For this case, neural coherence occurs via cooperation of regular firings of suprathreshold neurons. Unlike the suprathreshold case, subthreshold neurons (below the threshold) cannot fire spontaneously without noise; they can fire only with the help of noise. Stochastic excitatory coherence (i.e., collective coherence between noise-induced spikings by synaptic excitation) was observed in a population of excitatory subthreshold neurons \cite{CR,Kim}. Due to the stochastic excitatory coherence, synaptic current, injected into each individual neuron, becomes temporally coherent. Hence, temporal coherence resonance of an individual subthreshold neuron in the network may be enhanced. Furthermore, stochastic inhibitory coherence (i.e., collective coherence between noise-induced spikings by synaptic inhibition) was also investigated in a population of inhibitory subthreshold ML neurons exhibiting the type-II excitability \cite{Lim}. Weak stochastic inhibitory coherence was thus found to appear via cooperation of individual irregular oscillations (i.e., a regular small-amplitude ensemble-averaged oscillation emerges from sparsely synchronized neurons discharging irregularly at lower rates than the network oscillation). These sparsely synchronized neural oscillations have been intensively investigated in other types of neural networks \cite{Brunel} and they are believed to be associated with cortical rhythms in cognition  [e.g. ultrafast rhythm (100-200 Hz), gamma rhythm (30-100 Hz) and beta rhythm (15-30 Hz)] with irregular and sparse neural discharges \cite{W_Review,Brunel}.

In contrast to the case of subthreshold type-II ML neurons, no stochastic inhibitory coherence is observed in a homogeneous population of subthreshold type-I ML neurons. Hence, subthreshold type-I integrator neurons (without subthreshold oscillations) seem to be much more difficult to synchronize by inhibition than subthreshold type-II resonator neurons (exhibiting subthreshold oscillations). To take into consideration the effect of (spontaneously firing) suprathreshold neurons on the inhibitory coherence, we consider a heterogeneous inhibitory ensemble of subthreshold and suprathreshold type-I ML neurons.
Heterogeneity (or diversity) has been found to make constructive effects on collective coherence in various physical, biological, neural, and social systems \cite{Diversity}. In Section \ref{sec:IC}, we investigate  heterogeneity-induced inhibitory coherence by increasing the fraction of suprathreshold neurons $P_{supra}$ in the whole population. As $P_{supra}$ passes a threshold value $P_{supra}^*$ suprathreshold neurons begin to synchronize and they play the role of coherent inhibitors for the emergence of inhibitory coherence in the whole heterogeneous population. Thus, for $P_{supra}>P_{supra}^*$ the ensemble-averaged global potential $V_G$ exhibits a regular small-amplitude oscillation. For this coherent case, individual suprathreshold neurons exhibit intermittent spikings phase-locked to $V_G$ at random multiples of the period of $V_G$. Due to the stochastic spike skipping of suprathreshold neurons, the interspike interval (ISI) histogram has multiple peaks and partial occupation occurs in the raster plot of neural spikes. In addition to the coherent intermittent spikings, coherent subthreshold (small-amplitude) hopping oscillations also appear in the individual potentials of suprathreshold neurons. Thus, sparsely synchronized suprathreshold neurons exhibit coherent mixed-mode oscillations with two well-separated frequencies, a fast subthreshold hopping frequency imposed by the collective network oscillation and a lower firing frequency of individual suprathreshold neurons. By virtue of their coherent inhibition sparsely synchronized suprathreshold neurons suppress noisy activities of subthreshold neurons. Thus, only coherent fast subthreshold hopping oscillations (without spikings) appear in the individual potentials of subthreshold neurons. We also characterize this heterogeneity-induced inhibitory coherence in terms of ``statistical-mechanical'' spike-based and correlation-based measures, and find that the degree of inhibitory coherence increases as $P_{supra}$ is increased from $P_{supra}^*$. In a real brain, each neuron is coupled to only a certain number of neurons which is much smaller than the total number of neurons. The effect of sparseness of synaptic connectivity on the inhibitory coherence is briefly discussed by varying the average number of synaptic inputs per neuron $M_{syn}$ in a random network. Emergence of inhibitory coherence is thus found to persist until $M_{syn}$ is larger than a threshold value $M_{syn}^*$. We also confirm the universality of the heterogeneity-induced inhibitory coherence in a population of canonical type-I quadratic integrate-and-fire neurons \cite{QIF}. This kind of heterogeneity-induced weak inhibitory coherence might be associated with cortical rhythms with stochastic and sparse neural discharges which contribute to cognitive functions in the cerebral cortex (e.g., information integration, working memory, and selective attention) \cite{W_Review,Brunel}. Finally, a summary is given in Section \ref{sec:SUM}.

\section{Heterogeneous Ensemble of Inhibitory Type-I ML Neurons}
\label{sec:HE}

In this section we describe the biological neuron model used in our computational study. We consider a heterogeneous inhibitory ensemble of $N$ globally-coupled subthreshold and suprathreshold type-I neurons. As an element in our neural system, we choose the conductance-based ML neuron model, originally proposed to describe the time-evolution pattern of the membrane potential for the giant muscle fibers of barnacles \cite{ML1}. The population dynamics in this neural network is governed by the following set of differential equations:
\begin{subequations}
\begin{eqnarray}
C \frac{dv_i}{dt} &=& -I_{ion,i}+I_{DC,i} +D \xi_{i} -I_{syn,i}, \\
\frac{dw_i}{dt} &=& \phi \frac{(w_{\infty}(v_i) -
w_i)}{\tau_R(v_i)}, \\
\frac{ds_i}{dt}&=& \alpha s_{\infty}(v_i) (1-s_i) -
\beta s_i, \;\;\; i=1, \cdots, N,
\end{eqnarray}
\label{eq:CML1}
\end{subequations}
where
\begin{subequations}
\begin{eqnarray}
I_{ion,i} &=& I_{Ca,i} + I_{K,i} + I_{L,i} \\
&=& g_{Ca} m_{\infty} (v_i) (v_i - V_{Ca}) + g_K w_i (v_i - V_K) \nonumber \\
  && + g_L (v_i - V_L), \\
I_{syn,i} &=& \frac{J}{N-1} \sum_{j(\ne i)}^N s_j(t) (v_i - V_{syn}), \label{eq:Syn} \\
m_{\infty}(v) &=& 0.5 \left[ 1+\tanh \left\{ (v-V_1)/{V_2}
\right\} \right], \\
 w_{\infty}(v)&=& 0.5 \left[ 1+\tanh \left\{
(v-V_3)/{V_4} \right\} \right], \\
\tau_{R}(v) &=& 1/ \cosh \left\{ (v-V_3)/(2V_4) \right\}, \\
s_{\infty} (v_i) &=& 1/[1+e^{-(v_i-v_t)/\delta}].
 \end{eqnarray}
\label{eq:CML2}
\end{subequations}
Here, the state of the $i$th neuron at a time $t$ (measured in
units of ms) is characterized by three state variables: the
membrane potential $v_i$ (measured in units of mV), the slow
recovery variable $w_i$ representing the activation of the $K^+$
current (i.e., the fraction of open $K^+$ channels), and the
synaptic gate variable $s_i$ denoting the fraction of open
synaptic ion channels. In Eq.~(\ref{eq:CML1}a), $C$ represents the
capacitance of the membrane of each neuron, and the time evolution
of $v_i$ is governed by four kinds of source currents.

The total ionic current $I_{ion,i}$ of the $i$th neuron consists
of the calcium current $I_{Ca,i}$, the potassium current
$I_{K,i}$, and the leakage current $I_{L,i}$. Each ionic current
obeys Ohm's law. The constants $g_{Ca}$, $g_{K}$, and $g_{L}$
are the maximum conductances for the ion and the leakage channels,
and the constants $V_{Ca}$, $V_K$, and $V_L$ are the reversal
potentials at which each current is balanced by the ionic
concentration difference across the membrane. Since the calcium
current $I_{Ca,i}$ changes much faster than the potassium current
$I_{K,i}$, the gate variable $m_i$ for the $Ca^{2+}$ channel is
assumed to always take its saturation value $m_\infty(v_i)$. On
the other hand, the activation variable $w_i$ for the $K^{+}$
channel approaches its saturation value $w_{\infty}(v_i)$ with a
relaxation time $\tau_R(v_i) / \phi$, where $\tau_R$ has a
dimension of ms and $\phi$ is a (dimensionless) temperature-like
time scale factor.

Each ML neuron is also stimulated by a DC current $I_{DC,i}$ and
a Gaussian white noise $\xi_i$ [see the 2nd and 3rd terms in
Eq.~(\ref{eq:CML1}a)] satisfying $\langle \xi_i(t) \rangle =0$
and $\langle \xi_i(t)~\xi_j(t') \rangle = \delta_{ij}~\delta(t-t')$,
where $\langle\cdots\rangle$ denotes the ensemble average. The noise
$\xi_i$ is a parametric one which randomly perturbs the strength of
the applied current $I_{DC,i}$, and its intensity is controlled by
the parameter $D$. Depending on the system parameters, the ML neuron may exhibit either type-I or type-II excitability \cite{Ex2}. Throughout this
paper, we consider the case of type-I excitability where $g_{Ca}
= 4~ {\rm mS/cm^2},\, g_{K} = 8~ {\rm mS/cm^2},\, g_{L} = 2~
{\rm mS/cm^2},$ $V_{Ca} = 120~ {\rm mV},\, V_{K}=-84~ {\rm mV},\,
V_{L} = -60~ {\rm mV},$ $C = 20~ \mu {\rm F/cm^2},\, \phi = 1/15,$
$V_1 = -1.2~ {\rm mV},\, V_2 = 18~ {\rm mV},\, V_3 = 12~ {\rm mV},$
and $V_4 = 17.4~ {\rm mV}$. (For comparison, a result on the
order parameter is given for the type-II case where the values of
the above parameters are the same as those in the type-I case
except that $g_{Ca} = 4.4~ {\rm mS/cm^2}$, $\phi = 0.04$,
$V_3 = 2~ {\rm mV}$, and $V_4 = 30~ {\rm mV}$.)
For the type-I case, a transition from a resting state to a spiking
state occurs for $I_{DC}^*=40$ $\mu {\rm A /cm^2}$ via a saddle-node
bifurcation on an invariant circle \cite{Ex2}, and a firing begins
at arbitrarily low frequency. On the other hand, a type-II neuron
exhibits a jump from a resting state to a spiking state through
a subcritical Hopf bifurcation for $I_{DC,h}^*=93.9$
$\mu {\rm A /cm^2}$ by absorbing an unstable limit cycle born via
fold limit cycle bifurcation for $I_{DC,l}^*=88.3$
$\mu {\rm A /cm^2}$ \cite{Ex2}, and hence the firing frequency
begins from a non-zero value. Here a spread in the value of
the DC input current $I_{DC}$ is taken into consideration, and
thus for each subthreshold (suprathreshold) neuron a value of
$I_{DC,i}$ is randomly chosen with a uniform probability in the
range of $(I_{DC}^*-\Delta, I_{DC}^*)$ [$(I_{DC}^*, I_{DC}^* +\Delta)]$
for the type-I case and in the range of $(I_{DC,l}^*-\Delta, I_{DC,l}^*)$
[$(I_{DC,h}^*, I_{DC,h}^* +\Delta)]$ for the type-II case, where
the value of the spread parameter $\Delta$ is set as $\Delta=10$
$\mu {\rm A /cm^2}$.

We consider a heterogeneous inhibitory ensemble of $N$ globally-coupled subthreshold and suprathreshold ML neurons where the fraction of suprathreshold neurons is given by $P_{supra} = \frac {N_{supra}} {N}$ ($N_{supra}$: number of suprathreshold neurons). The last term in Eq.~(\ref{eq:CML1}a) represents
the synaptic coupling between neurons in the network. Each neuron is connected to all the other ones through global synaptic couplings. $I_{syn,i}$ of
Eq.~(\ref{eq:CML2}c) represents such synaptic current injected into
the $i$th neuron. Here the coupling strength is controlled by the
parameter $J$ and $V_{syn}$ is the synaptic reversal potential. We
use $V_{syn}=-80$ mV for the inhibitory synapse. The synaptic gate
variable $s$ obeys the 1st order kinetics of Eq.~(\ref{eq:CML1}c)
\cite{GR,WB}. Here the normalized concentration of synaptic
transmitters $s_{\infty}(v)$, activating the synapse, is assumed to
be an instantaneous sigmoidal function of the membrane potential with
a threshold $v_t$ in Eq.~(\ref{eq:CML2}g), where we set $v_t=0$ mV
and $\delta=2$ mV. The transmitter release occurs when the neuron emits a spike (i.e., its potential $v$ is larger than $v_t$). For the inhibitory GABAergic synapse (involving the $\rm {GABA_A}$ receptors), the synaptic channel opening rate, corresponding to the inverse of the synaptic rise time $\tau_r$, is $\alpha=10$ ${\rm ms}^{-1}$, and the synaptic closing rate $\beta$, which is the inverse of the synaptic decay time $\tau_d$, is $\beta=0.1$
${\rm ms}^{-1}$ \cite{BK}. Hence, $I_{syn}$ rises fast and decays
slowly.

Numerical integration of Eq.~(\ref{eq:CML1}) is done using the Heun
method \cite{SDE} (with the time step $\Delta t=0.01$ ms) similar to
the second-order Runge-Kutta method, and data for $(v_i,w_i,s_i)$
$(i=1,\dots,N)$ are obtained with the sampling time interval
$\Delta t=1$ ms. For each realization of the stochastic process in
Eq.~(\ref{eq:CML1}), we choose a random initial point
$[v_i(0),w_i(0),s_i(0)]$ for the $i$th $(i=1,\dots, N)$ neuron
with uniform probability in the range of $v_i(0) \in (-70,50)$,
$w_i(0) \in (0.0,0.6)$, and $s_i(0) \in (0.0,1.0)$.

\section{Heterogeneity-Induced Inhibitory Coherence}
\label{sec:IC}

In this section, we are concerned about inhibitory coherence in a heterogeneous ensemble of $N$ globally-coupled subthreshold and suprathreshold type-I ML neurons. By increasing the fraction of suprathreshold neurons $P_{supra}$, we investigate the heterogeneity-induced inhibitory coherence. (Hereafter, for convenience we omit the dimensions of $I_{DC}$, $D$, and $J$.)

We first consider a homogeneous population (corresponding to the case of $P_{supra}=0$) composed of only subthreshold type-I ML neurons, and study the inhibitory coherence by varying both the coupling strength $J$ and the noise intensity $D$. Emergence of inhibitory coherence may be well described by the (ensemble-averaged) global potential,
\begin{equation}
 V_G (t) = \frac {1} {N} \sum_{i=1}^{N} v_i(t).
\label{eq:GPOT}
\end{equation}
In the thermodynamic limit $(N \rightarrow \infty)$, a collective state becomes coherent if $\Delta V_G(t)$ $(= V_G(t) - \overline{V_G(t)})$ is non-stationary (i.e., an oscillating global potential $V_G$ appears for a coherent case), where the overbar represents the time average. Otherwise (i.e., when $\Delta V_G$ is stationary), it becomes incoherent. Thus, the mean square deviation of the global potential $V_G$ (i.e., time-averaged fluctuations of $V_G$),
\begin{equation}
{\cal{O}} \equiv \overline{(V_G(t) - \overline{V_G(t)})^2},
 \label{eq:Order}
\end{equation}
plays the role of an order parameter used for describing the coherence-incoherence transition \cite{Order}. For the coherent (incoherent) state, the order parameter $\cal{O}$ approaches a non-zero (zero) limit value as $N$ goes to the infinity.
\begin{figure}
\includegraphics[width=\columnwidth]{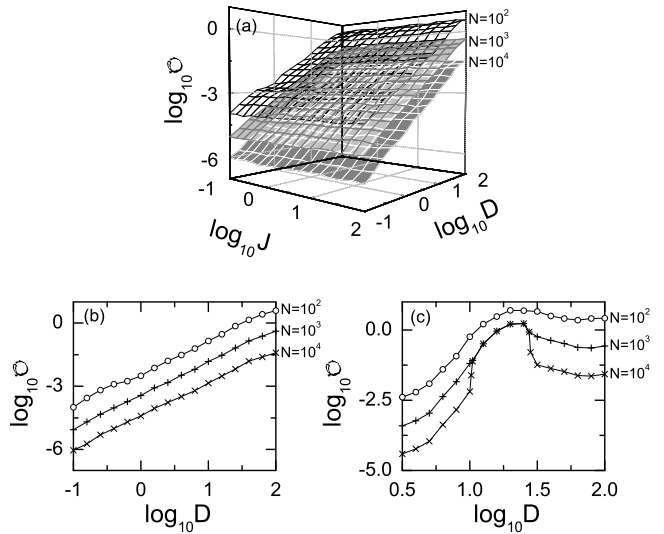}
\caption{Plots of the order parameter $\cal{O}$ versus (a) both the
coupling strength $J$ and the noise intensity $D$ and versus (b) $D$
for $J=20$ ${\rm mS /cm^2}$ in $N$ globally-coupled inhibitory
subthreshold type-I ML neurons. The value of $I_{DC,i}$ for each subthreshold
type-I neuron is randomly chosen with a uniform probability in the range
of $(I_{DC}^*-\Delta, I_{DC}^*)$ where $I_{DC}^*=40$ $\mu {\rm A /cm^2}$
and $\Delta=10$ $\mu {\rm A /cm^2}$. (c) Plots of $\cal{O}$ versus $D$ for
$J=3$ ${\rm mS /cm^2}$ in $N$ globally coupled inhibitory subthreshold
type-II ML neurons. The value of $I_{DC,i}$ for each subthreshold type-II
neuron is randomly chosen with a uniform probability in the range of
$(I_{DC,l}^*-\Delta, I_{DC,l}^*)$ where $I_{DC,l}^*=88.3$
$\mu {\rm A /cm^2}$ and $\Delta=10$ $\mu {\rm A /cm^2}$.
}
\label{fig:ORDER}
\end{figure}
Figure \ref{fig:ORDER}(a) shows plots of the order parameter versus both the coupling strength $J$ and the noise intensity $D$. As $N$ is increased, the order parameter tends to decrease, independently of $J$ and $D$. An example of the order parameter is shown in Fig.~\ref{fig:ORDER}(b) for $J=20$. For any given $D$, $\cal O$ is found to decrease as $N$ is increased. Hence, only incoherent states exist, irrespectively of $D$. This is in contrast to the case of subthreshold type-II neurons exhibiting inhibitory coherence. Figure \ref{fig:ORDER}(c) shows plots of the order parameter versus the noise intensity for the type-II case of $J=3$. Unlike the type-I case coherent states exist in an intermediate range of noise intensity [$D_l^*(\simeq 10.3) < D < D_h^*(\simeq 27.9)$] where the order parameter approaches a non-zero limit value as $N$ increases.

To take into consideration the effect of (spontaneously firing) suprathreshold neurons on the inhibitory coherence, we consider a heterogeneous population consisting of subthreshold and suprathreshold type-I ML neurons for $J=20$. For convenience, we set the value of noise intensity as $D=8$ and investigate the heterogeneity effect on the inhibitory coherence by increasing the fraction of suprathreshold neurons $P_{supra}$.
\begin{figure}
\includegraphics[width=\columnwidth]{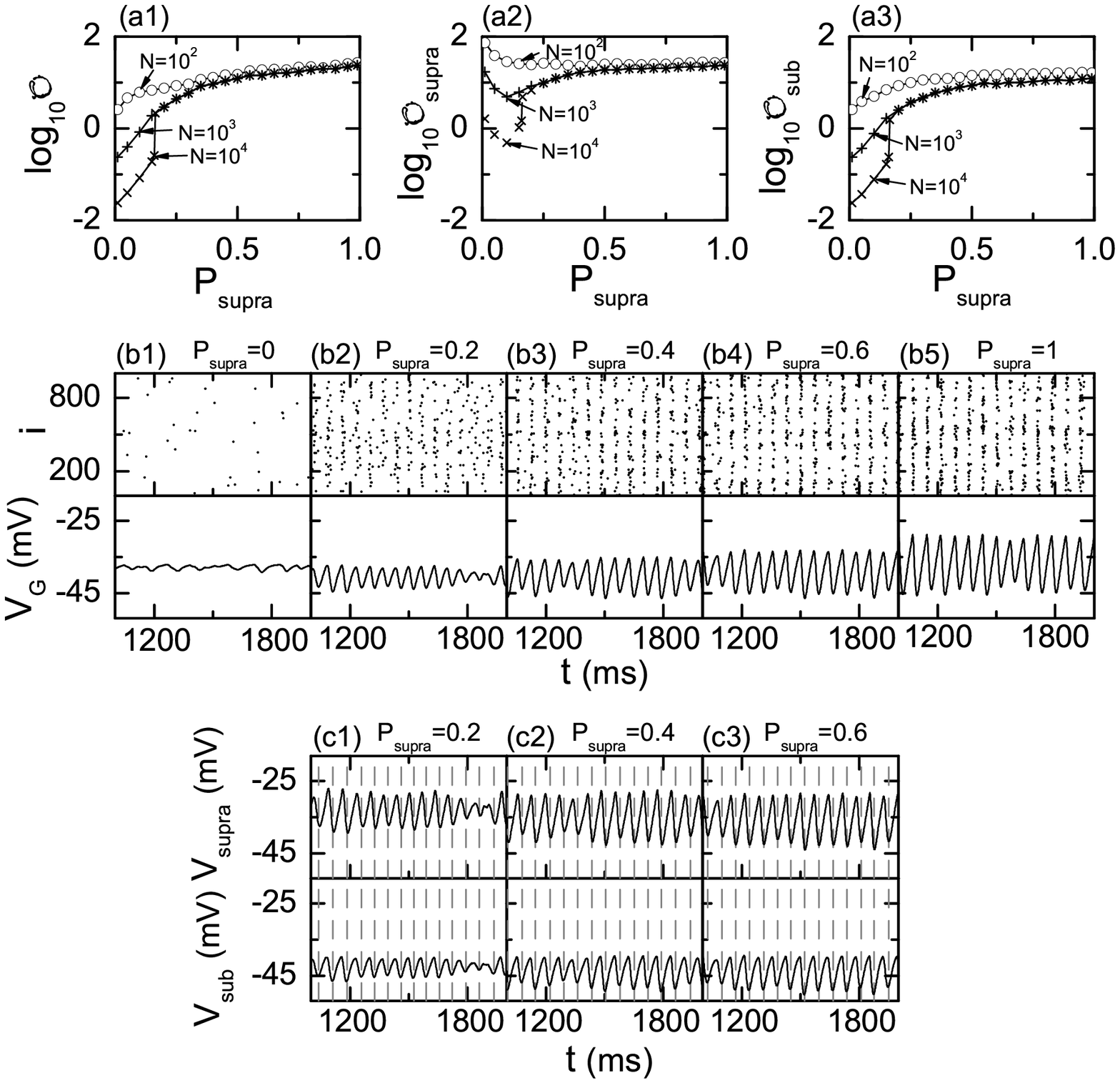}
\caption{Order parameters, raster plots of neural spikes, and time series of global potentials in the heterogeneous ensemble of $N$
globally-coupled inhibitory type-I ML neurons for $J=20$
${\rm mS /cm^2}$ and $D=8$ ${\rm \mu A \cdot {ms}^{1/2}/cm^2}$; $N=10^3$
in (b1)-(b5) and (c1)-(c3). The value of $I_{DC,i}$ for each subthreshold (suprathreshold) type-I ML neuron is randomly chosen with a uniform probability in the range of $(I_{DC}^*-\Delta, I_{DC}^*)$ [$(I_{DC}^*, I_{DC}^* +\Delta)]$ where $I_{DC}^*=40$ $\mu {\rm A /cm^2}$ and $\Delta=10$ $\mu {\rm A /cm^2}$.
Plots of the order parameter $\cal{O}$ versus the fraction of suprathreshold neurons $P_{supra}$ in (a1) the whole population and in the two subpopulations of (a2) the suprathreshold and (a3) the subthreshold neurons. Raster plots and time series of the global potential $V_G$ in the whole population for $P_{supra}=$ (b1) 0, (b2) 0.2, (b3) 0.4, (b4) 0.6, and (b5) 1.0. Time series of the
subensemble-averaged potentials $V_{supra}$ and $V_{sub}$ in the two
subpopulations of the suprathreshold and the subthreshold neurons
for $P_{supra}=$ (c1) 0.2, (c2) 0.4, and (x
c3) 0.6. Vertical dashed
lines in (c1)-(c3) represent the times at which local minima of $V_G$ appear.}
\label{fig:CT}
\end{figure}
Figure \ref{fig:CT}(a1) shows plots of the order parameter $\cal{O}$ versus $P_{supra}$ in the whole population. As $P_{supra}$ passes a threshold value $P_{supra}^* (\simeq 0.16)$, a transition from an incoherent to a coherent state occurs. As shown in Fig.~\ref{fig:CT}(a1), it is enough to consider only the case of $N=10^3$ for the study of inhibitory coherence because the order parameter $\cal{O}$ becomes saturated for $N=10^3$. Hereafter we set the number of neurons as $N=10^3$ in all cases except the calculation of the order parameter. For an incoherent case of $P_{supra}=0$, the raster plot consists of randomly scattered sparse spikes and the global potential $V_G$ exhibits nearly a stationary irregular oscillation [see Fig.~\ref{fig:CT}(b1)]; the amplitude of $V_G$ decreases with further increase in $N$. However, when passing the threshold $P_{supra}^*$ partially-occupied ``stripes'' (composed of spikes and indicating collective coherence) appear in the raster plot together with regularly-oscillating small-amplitude $V_G$ with frequency $f_G$ $(=13.8$ Hz)
[see Fig.~\ref{fig:CT}(b2)]. As $P_{supra}$ is further increased, both the pacing degree of spikes and the amplitude of $V_G$ (representing the degree of collective coherence) increase, as shown in Figs.~\ref{fig:CT}(b3)-\ref{fig:CT}(b5) where $f_G=$ (b3) 14.3 Hz, (b4) 13.6 Hz, and (b5) 14.2 Hz. This kind of weak inhibitory coherence also occurs in each subpopulation of the subthreshold and the suprathreshold neurons. As in the case of the whole population, emergence of inhibitory coherence in the subpopulations may be well described by the subensemble-averaged potentials $V_{supra}$ and $V_{sub}$,
\begin{subequations}
\begin{eqnarray}
 V_{supra}(t) &=& \frac {1} {N_{supra}} \sum_{i=1}^{N_{supra}} v_i(t), \\
 V_{sub}(t) &=& \frac {1} {N_{sub}} \sum_{i=1}^{N_{sub}} v_i(t),
\label{eq:SubPOT}
\end{eqnarray}
\end{subequations}
where $N_{supra}$ ($N_{sub}$) is the number of suprathreshold
(subthreshold) neurons. Then the order parameters ${\cal O}_{supra}$
and ${\cal O}_{sub}$, defined by the mean square deviation of
$V_{supra}$ and $V_{sub}$,
\begin{subequations}
\begin{eqnarray}
{\cal O}_{supra} &\equiv& \overline{(V_{supra}(t) -
\overline{V_{supra}(t)})^2},\\
{\cal O}_{sub} &\equiv& \overline{(V_{sub}(t) -
\overline{V_{sub}(t)})^2},
\label{eq:SubOrder}
\end{eqnarray}
\end{subequations}
may be used for describing the coherence-incoherence transitions in the
subpopulations of suprathreshold and subthreshold neurons, respectively.
Plots of ${\cal O}_{supra}$ and ${\cal O}_{sub}$ versus $P_{supra}$
are shown in Figs.~\ref{fig:CT}(a2) and \ref{fig:CT}(a3), respectively.
Coherent transition in each subpopulation occurs at the same threshold
value $P_{supra}^*(\simeq 0.16)$. For the case of coherent states, not
only $V_{supra}$ but also $V_{sub}$ exhibits regular oscillations whose
amplitudes increase as $P_{supra}$ is increased [see
Figs.~\ref{fig:CT}(c1)-\ref{fig:CT}(c3) where vertical dashed lines in
$V_{supra}$ and $V_{sub}$ denote the times at which local minima of
$V_G$ appear]. Both $V_{supra}$ and $V_{sub}$ are phase-locked to $V_G$.

To further understand the emergence of inhibitory coherence, we examine the individual and the global output signals in the subpopulations of subthreshold and suprathreshold neurons.
\begin{figure}
\includegraphics[width=\columnwidth]{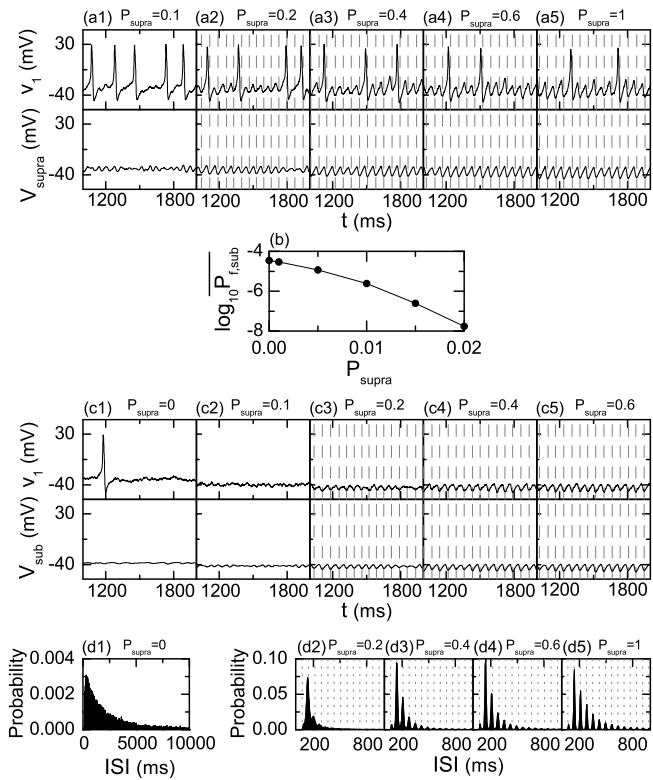}
\caption{Time series of the individual and the global potentials, the
average firing probability, and the interspike interval (ISI)
histogram in the heterogeneous ensemble of $N (=10^3)$
globally-coupled inhibitory type-I ML neurons for $J=20$ ${\rm mS /cm^2}$ and
$D=8$ ${\rm \mu A \cdot {ms}^{1/2}/cm^2}$. The value of $I_{DC,i}$
for each subthreshold (suprathreshold) type-I ML neuron is randomly
chosen with a uniform probability in the range of $(I_{DC}^*-\Delta,
I_{DC}^*)$ [$(I_{DC}^*, I_{DC}^* +\Delta)]$ where $I_{DC}^*=40$
$\mu {\rm A /cm^2}$ and $\Delta=10$ $\mu {\rm A /cm^2}$.
The individual potential $v_1$ of the first neuron and the
global potential $V_{supra}$ in the subpopulation of
suprathreshold neurons for $P_{supra}=$ (a1) 0.1, (a2)
0.2, (a3) 0.4, (a4) 0.6, and (a5) 1.0. Vertical dashed lines in
(a2)-(a5) represent the times at which local minima of $V_{supra}$
appear. (b) Plot of the average firing probability $\overline{P_{f,sub}}$
versus the fraction of suprathreshold neurons $P_{supra}$. The individual potential $v_1$ of the first neuron and
the global potential $V_{sub}$ in the subpopulation of subthreshold
neurons for $P_{supra}=$ (c1) 0, (c2) 0.1, (c3) 0.2, (c4) 0.4,
and (c5) 0.6. Vertical dashed lines in (c3)-(c5) denote the times at
which local minima of $V_{sub}$ appear. ISI histograms in the whole
population for $P_{supra}$ (d1) 0, (d2) 0.2, (d3) 0.4, (d4) 0.6,
and (d5) 1.0; each ISI histogram is composed of $5 \times 10^4$ ISIs and
the bin size for the histogram is 5 ms. Vertical dotted lines in
(d2)-(d5) denote integer multiples of $T_G$ (period of $V_G$).}
\label{fig:COH}
\end{figure}
Figures \ref{fig:COH}(a1)-\ref{fig:COH}(a5) show the time series of the individual potential $v_1$ of the first neuron and the time series of the global potential $V_{supra}$ in the subpopulation of the suprathreshold neurons. $V_{supra}$ exhibits a regular small-amplitude oscillation for a coherent case, while it shows a nearly stationary irregular oscillation for an incoherent case. For the case of coherent states, individual suprathreshold neurons exhibit intermittent spikings phase-locked to $V_{supra}$ at random multiples of the period of $V_{supra}$ [see Figs.~\ref{fig:COH}(a2)-\ref{fig:COH}(a5) where dashed lines denote the times at which local minima of $V_{supra}$ appear]. This ``stochastic phase locking'' leading to stochastic spike skipping is well shown in the ISI histogram with multiple peaks [see Figs.~\ref{fig:COH}(d2)-\ref{fig:COH}(d5)]. In addition to these coherent intermittent spiking phases, coherent subthreshold (small-amplitude) hopping oscillations also appear in the individual potentials of suprathreshold neurons. Thus, suprathreshold neurons exhibit coherent mixed-mode oscillations with two well-separated frequency scales, a fast subthreshold hopping frequency $f_h$  imposed by the collective network oscillation with frequency $f_G$ $(\simeq$ 14 Hz) and a lower spiking frequency $f_s$ of individual suprathreshold neurons; $f_s=$ (a2) 3.6 Hz, (a3) 2.8 Hz, (a4) 2.3 Hz, and (a5) 1.8 Hz. These sparsely synchronized suprathreshold neurons play the role of coherent inhibitors for the emergence of inhibitory coherence in the whole heterogeneous population, as shown below. On the other hand, for the incoherent case only stochastic intermittent spikings occur without any coherent hoppings, as shown in Fig.~\ref{fig:COH}(a1). Figure \ref{fig:COH}(b) shows the plot of the average firing probability $\overline{P_{f,sub}}$ versus $P_{supra}$ in the subpopulation of subthreshold neurons (i.e., time-averaged fraction of firing subthreshold neurons in the subpopulation of subthreshold neurons). Due to inhibition $\overline{P_{f,sub}}$ decreases dramatically with respect to $P_{supra}$. For $P_{supra}>0.02$, one can disregard spikings of subthreshold neurons because $\overline{P_{f,sub}} (\sim 10^{-8})$ becomes very small. The time series of the individual potential $v_1$ of the first neuron and the time series of the global potential $V_{sub}$ in the subpopulation of
the subthreshold neurons are shown in Fig.~\ref{fig:COH}(c1)-\ref{fig:COH}(c5). A regular oscillation with small amplitude occurs in $V_{sub}$ for a coherent case, while $V_{sub}$ exhibits a nearly stationary irregular oscillation for an incoherent case. For the case of coherent states, sparsely synchronized suprathreshold neurons suppress noisy activities of subthreshold neurons by virtue of their coherent inhibition, and then individual subthreshold neurons exhibit only coherent subthreshold hoppings (without spikings), in contrast to the suprathreshold case. Figures \ref{fig:COH}(d1)-\ref{fig:COH}(d5) show the ISI histograms in the whole population. (As shown above, spiking neurons in the whole population are just suprathreshold ones for $P_{supra}>0.02.$) The ISI histogram for $P_{supra}=0$ has a very long tail, and hence the average value
$\langle {\rm ISI} \rangle (\simeq 4926$ ms) of ISIs is very large. As
$P_{supra}$ passes the threshold $P_{supra}^*$, multiple peaks tend to
appear at integer multiples of $T_G$ (period of $V_G$) [i.e., $n\,T_G$ $(n=1,2,3, \dots)$] [e.g., see the ISI histogram for $P_{supra}=0.2$; vertical dotted lines in the histogram denote integer multiples of $T_G (= 72.4$ ms)]. As $P_{supra}$ is further increased, ISI histograms with more distinct multiple peaks appear due to the stochastic spike skipping of the suprathreshold neurons, as shown in Figs.~\ref{fig:COH}(d3)-\ref{fig:COH}(d5) where $T_G$ = (d3) 69.9 ms, (d4) 73.3 ms, and (d5) 70.3 ms. The most probable peak appears at $2\,T_G$, and hence suprathreshold neurons fire mostly in alternate global cycles.

We characterize the heterogeneity-induced inhibitory coherence in terms of two kinds of ``statistical-mechanical'' spike-based and correlation-based measures. As shown in Figs.~\ref{fig:CT}(b1)-\ref{fig:CT}(b5), inhibitory coherence
may be well visualized in the raster plot of spikes. For a coherent case,
the raster plot is composed of partially-occupied stripes (indicating
collective coherence). To measure the degree of the collective coherence
seen in the raster plot, a new spike-based measure $M_s$ was introduced
by considering the occupation pattern and the pacing pattern of neural spikes
in the ``stripes'' \cite{Lim}. Particularly, the pacing degree between
spikes is determined in a statistical-mechanical way by quantifying
the average contribution of microscopic individual spikes to the global
potential $V_G$. The spiking coherence measure $M_i$ of the $i$th stripe is
defined by the product of the occupation degree $O_i$ of spikes (representing
the density of the $i$th stripe) and the pacing degree $P_i$ of spikes
(denoting the smearing of the $i$th stripe):
\begin{equation}
M_i = O_i \cdot P_i.
\label{eq:SM}
\end{equation}
The occupation degree $O_i$ in the $i$th stripe is given by the fraction of
spiking neurons:
\begin{equation}
   O_i = \frac {N_i^{(s)}} {N},
\end{equation}
where $N_i^{(s)}$ is the number of spiking neurons in the $i$th stripe.
For the full occupation, $O_i=1$, while for the partial occupation $O_i<1$.
The pacing degree $P_i$ of each microscopic spike in the $i$th stripe can be
determined in a statistical-mechanical way by taking into consideration its
contribution to the macroscopic global potential $V_G$. Each global cycle of
$V_G$ begins from its left minimum, passes the central maximum, and ends at the right minimum; the central maxima coincide with centers of stripes in the raster plot [see Figs.~\ref{fig:CT}(b2)-\ref{fig:CT}(b5)]. An instantaneous global phase $\Phi(t)$ of $V_G$ is introduced via linear interpolation in the two successive subregions forming a global cycle. The global phase $\Phi(t)$ between the left minimum (corresponding to the beginning point of the $i$th global cycle) and the central maximum is given by:
\begin{eqnarray}
&&\Phi(t) = 2\pi(i-3/2) + \pi \left(
\frac{t-t_i^{(min)}}{t_i^{(max)}-t_i^{(min)}} \right) \nonumber \\
&& {\rm for~~} t_i^{(min)} \leq  t < t_i^{(max)}
~~(i=1,2,3,\dots),
\end{eqnarray}
and $\Phi(t)$ between the central maximum and the right minimum (corresponding
to the beginning point of the $(i+1)$th cycle) is given by:
\begin{eqnarray}
&& \Phi(t) = 2\pi(i-1) + \pi \left(
\frac{t-t_i^{(max)}}{t_{i+1}^{(min)}-t_i^{(max)}} \right) \nonumber \\
&& {\rm for~~} t_i^{(max)} \leq  t < t_{i+1}^{(min)}
~~(i=1,2,3,\dots),
\end{eqnarray}
where $t_i^{(min)}$ is the beginning time of the $i$th global cycle (i.e., the time at which the left minimum of $V_G$ appears in the $i$th global cycle) and $t_i^{(max)}$ is the time at which the maximum of $V_G$ appears in the $i$th global cycle. Then, the contribution of the $k$th microscopic spike in the $i$th stripe occurring at the time $t_k^{(s)}$ to $V_G$ is given by $\cos \Phi_k$, where $\Phi_k$ is the global phase at the $k$th spiking time [i.e., $\Phi_k \equiv \Phi(t_k^{(s)})$]. A microscopic spike makes the most constructive (in-phase) contribution to $V_G$ when the corresponding
global phase $\Phi_k$ is $2 \pi n$ ($n=0,1,2, \dots$), while it makes the most
destructive (anti-phase) contribution to $V_G$ when $\Phi_i$ is $2 \pi (n-1/2)$. By averaging the contributions of all microscopic spikes in the $i$th stripe to $V_G$, we obtain the pacing degree of spikes in the $i$th stripe,
\begin{equation}
 P_i ={ \frac {1} {S_i}} \sum_{k=1}^{S_i} \cos \Phi_k,
\label{eq:PACING}
\end{equation}
where $S_i$ is the total number of microscopic spikes in the $i$th stripe.
By averaging $M_i$ of Eq.~(\ref{eq:SM}) over a sufficiently large number $N_s$ of stripes, we obtain the spike-based coherence measure $M_s$:
\begin{equation}
M_s =  {\frac {1} {N_s}} \sum_{i=1}^{N_s} M_i.
\label{eq:CM}
\end{equation}
\begin{figure}
\includegraphics[width=\columnwidth]{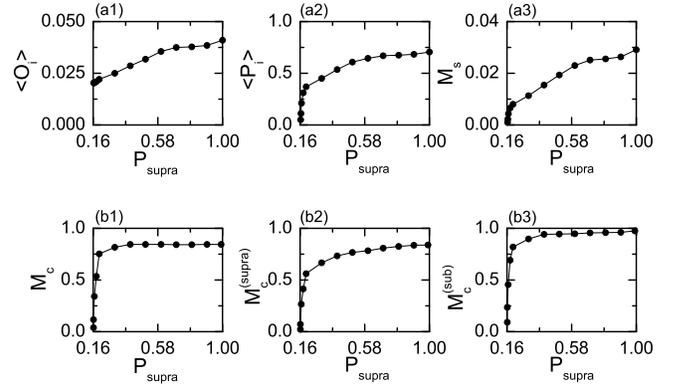}
\caption{``Statistical-mechanical'' coherence measures in the
heterogeneous ensemble of $N (=10^3)$ globally-coupled inhibitory
type-I ML neurons for $J=20$ ${\rm mS /cm^2}$ and $D=8$
${\rm \mu A \cdot {ms}^{1/2}/cm^2}$. (a) Spike-based coherence measure $M_s$:
(a1) plot of the average occupation degree $\langle O_i \rangle$
versus the fraction of suprathreshold neurons $P_{supra}$, (a2) plot of the average pacing degree $\langle P_i \rangle$ versus $P_{supra}$, and (a3) plot of the spiking coherence measure $M_s$ versus $P_{supra}$. To obtain
$\langle O_i \rangle$, $\langle P_i \rangle$, and $M_s$, we
follow the $3 \times 10^3$ stripes for each $P_{supra}$.
(b) Correlation-based coherence measure $M_c$: (b1) plot of $M_c$
versus $P_{supra}$ in the whole population and plots of $M_c^{(supra)}$
and $M_c^{(sub)}$ versus $P_{supra}$ in the two subpopulations
of (b2) suprathreshold and (b3) subthreshold neurons. The number of data
used for the calculation of a cross-correlation function
for each $P_{supra}$ is $2^{12}$.}
\label{fig:CM}
\end{figure}
By varying $P_{supra}$, we follow $3 \times 10^3$ stripes and measure the degree of collective spiking coherence in terms of $\langle O_i \rangle$ (average occupation degree), $\langle P_i \rangle$ (average pacing degree), and $M_s$ for 13 values of $P_{supra}$ in the coherent regime, and the results are shown in Figs.~\ref{fig:CM}(a1)-\ref{fig:CM}(a3). As $P_{supra}$ is increased, the average occupation degree $\langle O_i \rangle$ (denoting the average density of stripes in the raster plot) increases slowly, but its values are
very small ($\langle O_i \rangle <0.05$); only a fraction (less than 1/20) of the total neurons fire in each stripe [see Figs.~\ref{fig:CT}(b2)-\ref{fig:CT}(b5)]. This partial occupation results from stochastic spike skipping of individual neurons seen well in the multi-peaked ISI histograms [see Figs.~\ref{fig:COH}(d2)-\ref{fig:COH}(d5)].
On the other hand, the average pacing degree $\langle P_i \rangle$ increases rapidly near the threshold $P_{supra}^*$, and then it grows slowly. This tendency may be understood from the change in the structure of the ISI histograms. As $P_{supra}$ is increased, clear well-separated multiple peaks appear, and hence the average pacing degree of the stripes becomes better with increasing $P_{supra}$. In most region of the coherent region, the values of $\langle P_i \rangle$ are large in contrast to $\langle O_i \rangle$. However, the spiking measure $M_s$ of Eq.~(\ref{eq:CM}) (representing the collective spiking coherence) is very low due to the partial occupation in the raster plot.

For the coherent case, subthreshold neurons exhibit only coherent subthreshold hoppings without spikings. Hence, only suprathreshold neurons (exhibiting coherent intermittent spikings) make contribution to the spike-based measure $M_s$. From now on, we use another statistical-mechanical measure based on the ensemble average of cross-correlations between the global potential and the individual potentials \cite{Lim2}, and measure the degree of the inhibitory coherence in the subpopulations of the subthreshold and the suprathreshold neurons as well as in the whole population. The inhibitory coherence
in the whole population is quantified in terms of the coherence measure $M_c$ given by the ensemble average of the global-individual cross-correlations $C_i(0)$ between $V_G$ and $v_i$ at the zero-time lag:
\begin{equation}
M_c = \frac {1} {N} \sum_{i=1}^N C_i(0).
\label{eq:MC}
\end{equation}
Here the normalized cross-correlation function $C_i(\tau)$ between $V_G$ and $v_i$ is given by
\begin{equation}
C_i(\tau) = \frac {\overline{\Delta V_G(t + \tau)\, \Delta v_i(t)}}
                 {\sqrt{\overline{\Delta V_G^2(t)}}\, \sqrt{\overline{\Delta v_i^2(t)}}},
\label{eq:CCGL}
\end{equation}
where $\tau$ is the time lag, $\Delta V_G(t) = V_G(t) - \overline{V_G(t)}$,
$\Delta v_i(t) = v_i(t)- \overline{v_i(t)}$, and the overline denotes the time average. This correlation-based measure $M_c$ can be regarded as a ``statistical-mechanical'' measure because it quantifies the average contribution of (microscopic) individual potentials to the (macroscopic) global potential. Hence, $M_c$ is in contrast to the conventional microscopic measure based on the cross-correlations between the individual potentials. As in the case of the whole population, the degree of inhibitory coherence in each
subpopulation of the subthreshold and the suprathreshold neurons may be well quantified in terms of the coherence measures $M_c^{(sub)}$ and $M_c^{(supra)}$ based on the cross-correlations between the global potentials ($V_{sub}$ and $V_{supra}$) and the individual potentials,
\begin{subequations}
\begin{eqnarray}
M_c^{(supra)} &=& \frac {1} {N_{supra}} \sum_{i=1}^{N_{supra}} C_i^{(supra)}(0),\\
M_c^{(sub)} &=& \frac {1} {N_{sub}} \sum_{i=1}^{N_{sub}} C_i^{(sub)}(0),
\label{eq:MC_Sub}
\end{eqnarray}
\end{subequations}
where
\begin{subequations}
\begin{eqnarray}
C_i^{supra}(\tau) &=& \frac {\overline{\Delta V_{supra}(t + \tau)\, \Delta v_i(t)}}
                 {\sqrt{\overline{\Delta V_{supra}^2(t)}}\, \sqrt{\overline{\Delta v_i^2(t)}}}
                 \; (i=1,...,N_{supra}), \nonumber \\
                 && \\
C_i^{sub}(\tau) &=& \frac {\overline{\Delta V_{sub}(t + \tau)\, \Delta v_i(t)}}
                 {\sqrt{\overline{\Delta V_{sub}^2(t)}}\, \sqrt{\overline{\Delta v_i^2(t)}}}
                 \;(i=1,...,N_{sub}). \nonumber \\
                 &&
\label{eq:CCGL_Sub}
\end{eqnarray}
\end{subequations}
By varying $P_{supra}$, we measure the degree of inhibitory coherence in terms of the correlation-based measures $M_c$, $M_c^{(supra)}$, and $M_c^{(sub)}$ not only in the whole population, but also in the subpopulations of the subthreshold and the suprathreshold neurons, and the results are shown in Figs.~\ref{fig:CM}(b1)-\ref{fig:CM}(b3). All of the coherence measures increase rapidly near the threshold $P_{supra}^*$, and then they grow slowly. The values of these correlation-based measures are very large in contrast to the spiking coherence measure $M_s$. We also note that the degree of inhibitory coherence in the subpopulation of subthreshold neurons is higher than that
in the subpopulation of suprathreshold neurons. This can be understood from the
oscillating patterns of the global and the individual potentials. The global
potentials $V_{supra}$ and $V_{sub}$ exhibit small regular oscillations
[see Figs.~\ref{fig:COH}(a2)-\ref{fig:COH}(a5) and
Figs.~\ref{fig:COH}(c3)-\ref{fig:COH}(c5)]. 
Like the case of the global potential, the individual subthreshold neurons exhibit only coherent subthreshold hoppings, in contrast to the case of suprathreshold neurons exhibiting both the coherent intermittent spikings and the coherent hoppings. Hence, the cross-correlations between $V_{sub}$ and the individual potentials of subthreshold neurons become higher than those between $V_{supra}$ and the individual potentials of suprathreshold neurons.

\begin{figure}
\includegraphics[width=\columnwidth]{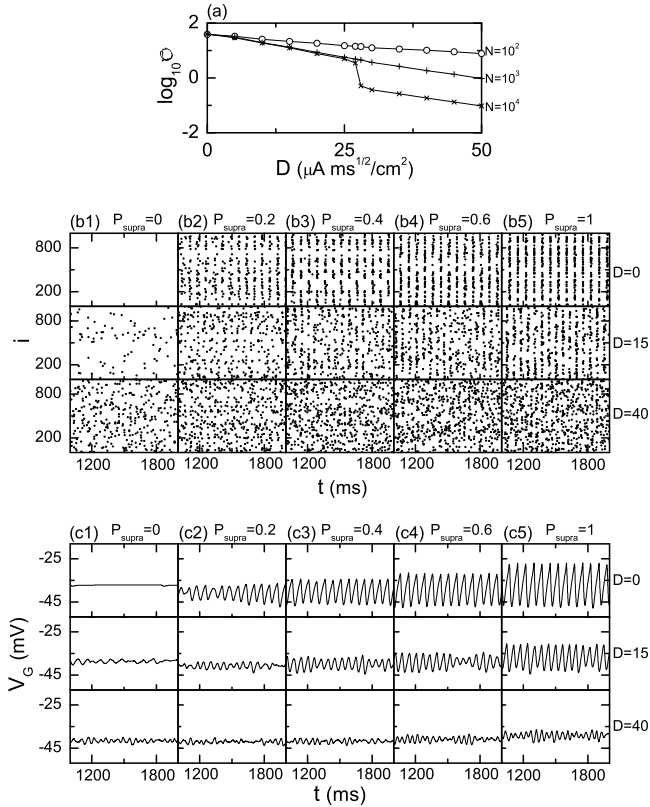}
\caption{(a) Plots of the order parameter $\cal{O}$ versus the noise intensity $D$ for $P_{supra}=1$, (b1)-(b5) raster plots of neural spikes and (c1)-(c5) global potentials $V_G$ for various values of $P_{supra}$ and $D$ in a heterogeneous ensemble of $N$ globally-coupled subthreshold and suprathreshold type-I ML neurons for $J=20$ ${\rm mS /cm^2}$; $N=10^3$ in (b1)-(b5) and (c1)-(c5). The value of $I_{DC,i}$ for each subthreshold (suprathreshold) type-I ML neuron is randomly chosen with a uniform probability in the range of $(I_{DC}^*-\Delta,
I_{DC}^*)$ [$(I_{DC}^*, I_{DC}^* +\Delta)]$ where $I_{DC}^*=40$
$\mu {\rm A /cm^2}$ and $\Delta=10$ $\mu {\rm A /cm^2}$.
}
\label{fig:Noise}
\end{figure}
In the above, we study the heterogeneity-induced inhibitory coherence for a fixed value of $D=8$ where $P_{supra}^* \simeq 0.16$. 
By varying the noise intensity $D$ we investigate the effect of noise on the inhibitory coherence for $J=20$. For $P_{supra}=1$, plots of the order parameter $\cal{O}$ versus $D$ are shown in Fig.~\ref{fig:Noise}(a). The degree of inhibitory coherence decreases monotonically with increasing $D$ from zero, and a transition to an incoherent state occurs when passing a threshold $D^*$ $(\simeq 28)$. Figures \ref{fig:Noise}(b1)-\ref{fig:Noise}(b5) and \ref{fig:Noise}(c1)-\ref{fig:Noise}(c5) show the raster plots of spikes and the global potentials $V_G$ for various values of $D$ and $P_{supra}$. For $P_{supra}=1$, with increasing $D$ the stripes in the raster plot become more smeared and the amplitude of $V_G$ decreases. Eventually when passing the threshold $D^*$, incoherent states appear (i.e., the raster plot consists of randomly scattered spikes and $V_G$ exhibits a nearly stationary irregular oscillation). Hence, as $D$ is increased the value of $P_{supra}^*$ (threshold value of the fraction of suprathreshold neurons for the emergence of inhibitory coherence) increases; $P_{supra}^*=$ 0.08 and 0.28 for $D=0$ and 15, respectively. Thus, for $D > D^*$ no inhibitory coherence emerges, as shown in the case of $D=40$.

\begin{figure}
\includegraphics[width=\columnwidth]{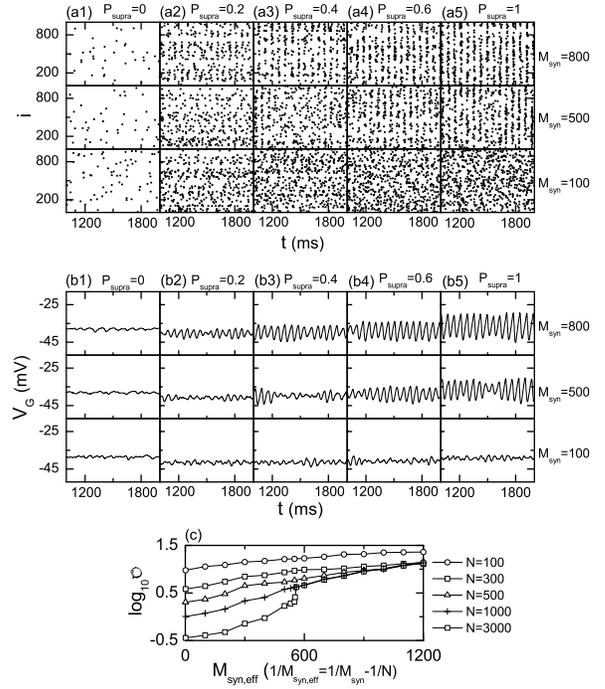}
\caption{(a1)-(a5) Raster plots of neural spikes and (b1)-(b5) global potentials $V_G$ for various values of $P_{supra}$ and $M_{syn}$ in a heterogeneous ensemble of $N$ $(=10^3)$ randomly-coupled subthreshold and suprathreshold type-I ML neurons for $J=20$ ${\rm mS /cm^2}$ and $D=8$ ${\rm \mu A \cdot {ms}^{1/2}/cm^2}$. (c) Plots of the order parameter $\cal{O}$ versus the effective average number of synaptic inputs per neuron $M_{syn,eff}$ $(1/M_{syn,eff}=1/M_{syn}-1/N)$ for $P_{supra}=1$. The value of $I_{DC,i}$ for each subthreshold (suprathreshold) type-I ML neuron is randomly
chosen with a uniform probability in the range of $(I_{DC}^*-\Delta,
I_{DC}^*)$ [$(I_{DC}^*, I_{DC}^* +\Delta)]$ where $I_{DC}^*=40$
$\mu {\rm A /cm^2}$ and $\Delta=10$ $\mu {\rm A /cm^2}$.
}
\label{fig:Random}
\end{figure}
So far, we consider the globally-coupled case. However, in a real brain each neuron is coupled to only a certain number of neurons which is much smaller than the total number of neurons. Due to the sparseness of the network architecture, the inhibitory coherence (seen in the globally-coupled case) is expected to be reduced or destroyed. It is often assumed in models that the coupling between neurons is random \cite{WB,BK,R1,R2,R3}. We briefly investigate the effect of sparse random connectivity on the inhibitory coherence for $J=20$ and $D=8$ by varying the average number of synaptic inputs per neuron $M_{syn}$ in a heterogeneous ensemble of $N$ randomly-coupled subthreshold and suprathreshold type-I ML neurons. Figs.~\ref{fig:Random}(a1)-\ref{fig:Random}(a5) and \ref{fig:Random}(b1)-\ref{fig:Random}(b5) show the raster plots of spikes and the global potentials $V_G$ for various values of $M_{syn}$ and $P_{supra}$  when $N=10^3$. For $P_{supra}=1$, with decreasing $M_{syn}$ the stripes of spikes in the raster plot become more smeared and the amplitude of $V_G$ decreases. Eventually, incoherent states appear when passing a threshold $M_{syn}^*$ (i.e., the raster plot is composed of randomly scattered spikes and $V_G$ shows a nearly stationary irregular oscillation). Hence, as $M_{syn}$ is decreased from $N-1$ (corresponding to the globally-coupled case), a larger fraction of suprathreshold neurons is necessary for the appearance of inhibitory coherence (e.g., see the cases of $M_{syn}=$800 and 500). Thus, for $M_{syn} < M_{syn}^*$ inhibitory coherence disappears, as shown in the case of $M_{syn}=100$. As is well known, $M_{syn}$ [rather than $P_{syn}$ (i.e., the connection probability per neuron)] plays an appropriate sparseness parameter for the coherent transition because there exists a fixed threshold value $M_{syn}^*$ for large $N$, independently of $N$ \cite{W_Review,R3}. (In contrast, the threshold value of $P_{syn}$ depends on $N$.) When $P_{supra}=1$, plots of the order parameter $\cal{O}$ versus the effective average number of synaptic inputs per neuron $M_{syn,eff}$ $(1/M_{syn,eff}=1/M_{syn} - 1/N)$ are shown in Fig.~\ref{fig:Random}(c) for $N=100, 300, 500, 1000,$ and 3000, where the correction term $(\sim 1/N)$ takes into account the finite network size effect \cite{W_Review,R3}. Inhibitory coherence emerges when $M_{syn,eff}$ is larger than a threshold $M_{syn,eff}^*$ $(\simeq 553)$; for the case of $N=10^3$, $M_{syn}^* \simeq 356$.

Finally, we examine the universality of the heterogeneity-induced inhibitory coherence in a population of $N$ globally-coupled subthreshold and suprathreshold quadratic integrate-and-fire (QIF) neurons \cite{QIF}. The QIF neuron model is canonical in the sense that any type-I excitable systems can be transformed into the form of the QIF model by a continuous change of variables. Here, we use a QIF model with parameters derived from the Wang-Buzsaki conductance-based model \cite{HM}. These QIF neurons interact via the same inhibitory GABAergic synapses as those in the case of the ML neurons. The dynamics of the membrane potential $v_i$ of each QIF neuron is governed by the following set of differential equations:
\begin{equation}
C \frac{dv_i}{dt} = A\, (v_i - v^*)^2 +I_i - I^* + D \xi_{i} -I_{syn,i},\;\;\; i=1, \cdots, N,
 \end{equation}
with the auxiliary after-spike resetting: if $v_i \geq v_t$ (threshold potential for spiking), $v_i \leftarrow v_r$ (resetting potential), where $t$ denotes the time (measured in units of ms), $C=0.9467~ \mu{\rm F/cm^2}$, $A=0.012875~{\rm mS/cm^2/mV}$, $v^*=-59.5462~{\rm mV}$, $I^*=0.1601~\mu{\rm A/cm^2}$, $v_t=-26.3462~{\rm mV}$, and $v_r=-64.1462~{\rm mV}$. The $i$th neuron is stimulated by the applied current $I_i$ and the white Gaussian noise $\xi_i$ whose intensity is controlled by the parameter $D$. The synaptic current injected into the $i$th neuron $I_{syn,i}$ is given in Eq.~(\ref{eq:Syn}) where the coupling strength is controlled by the parameter $J$, $V_{syn}(=-75~{\rm mV})$ is the synaptic reversal potential, and the synaptic gate variable $s$ obeys the 1st order kinetics of Eq.~(\ref{eq:CML1}c) along with Eq.~(\ref{eq:CML2}g) with $\delta=2~{\rm mV}$. For the single QIF neuron, a transition from a resting to a spiking state occurs when $I_i$ passes the rheobase $I^*$ via a saddle-node bifurcation on an invariant circle. Here a spread in the value of $I_i$ is taken into consideration, and thus for each subthreshold (suprathreshold) neuron a value of $I_i$ is randomly chosen with a uniform probability in the range of $(I^*-\Delta, I^*)$ [$(I^*, I^* +\Delta)]$, where the value of the spread parameter $\Delta$ is set as $\Delta=0.1~\mu{\rm A/cm^2}$. By increasing the fraction of suprathreshold neurons $P_{supra}$, we investigate the heterogeneity-induced inhibitory coherence for $J=5$ and $D=0.1$. 
\begin{figure}
\includegraphics[width=\columnwidth]{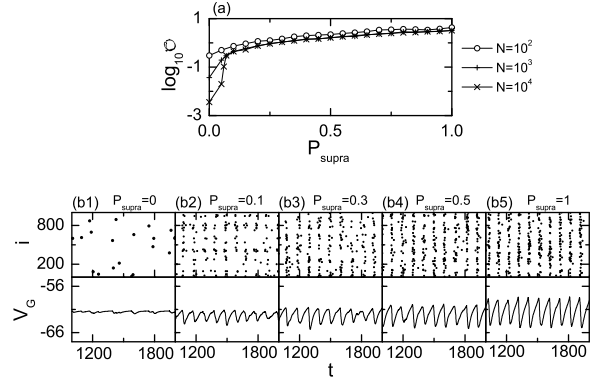}
\caption{Order parameters, raster plots of neural spikes, and time
series of global potentials in the heterogeneous ensemble of $N$
globally-coupled inhibitory QIF neurons for $J=5$ ${\rm mS /cm^2}$ and $D=0.1$ ${\rm \mu A \cdot {ms}^{1/2}/cm^2}$; $N=10^3$ in (b1)-(b5). The value of $I_{i}$ for each subthreshold (suprathreshold) QIF neuron is randomly chosen with a uniform probability in the range of $(I^*-\Delta, I^*)$ [$(I^*, I^* +\Delta)]$ where $I^*=0.1601$ $\mu {\rm A /cm^2}$ and $\Delta=0.1$ $\mu {\rm A /cm^2}$. Plots of the order parameter $\cal{O}$  versus the fraction of suprathreshold neurons $P_{supra}$ in (a). Raster plots of neural spikes and time series of the global potential $V_G$ for $P_{supra}=$ (b1) 0, (b2) 0.1, (b3) 0.3, (b4) 0.5 and (b5) 1.0.}
\label{fig:QIF}
\end{figure}
Figure \ref{fig:QIF}(a) shows plots of the order parameter $\cal{O}$ versus $P_{supra}$. As $P_{supra}$ passes a threshold value $P_{supra}^* (\simeq 0.05)$, a transition from an incoherent to a coherent state occurs like the case of ML neurons [see Fig.~\ref{fig:CT}(a1)]. For an incoherent case of $P_{supra}=0$, the raster plot consists of randomly scattered sparse spikes and the global potential $V_G$ exhibits a nearly stationary irregular oscillation, as shown in Fig.~\ref{fig:QIF}(b1). However, when passing the threshold $P_{supra}^*$ partially-occupied ``stripes'' appear in the raster plot along with regularly-oscillating small-amplitude global potential $V_G$ [see Fig.~\ref{fig:QIF}(b2) for $P_{supra}=0.1$]. As $P_{supra}$ is further increased, both the pacing degree of spikes and the amplitude of $V_G$ (representing the degree of inhibitory coherence) increase, as shown in Figs.~\ref{fig:QIF}(b3)-\ref{fig:QIF}(b5).

\section{Summary}
\label{sec:SUM}
We have studied the heterogeneity-induced inhibitory coherence by increasing the fraction of suprathreshold neurons $P_{supra}$ in an ensemble of globally-coupled subthreshold and suprathreshold type-I ML neurons. For $P_{supra}=0$ no inhibitory coherence has been observed, which implies that subthreshold type-I neurons are difficult to synchronize by synaptic inhibition. However, as $P_{supra}$ passes a threshold value $P_{supra}^*$, a coherent transition occurs in the subpopulation of suprathreshold neurons, and these synchronized suprathreshold neurons play the role of coherent inhibitors for the emergence of inhibitory coherence in the whole heterogeneous population. Consequently, for $P_{supra}>P_{supra}^*$ a regular population rhythm with small amplitude appears in the ensemble-averaged global potential $V_G$. For this coherent case, both the coherent intermittent spiking and the coherent subthreshold (small-amplitude) hopping phases appear in the individual potentials of suprathreshold neurons. Thus, these sparsely synchronized suprathreshold neurons exhibit coherent mixed-mode oscillations with two well-separated frequency scales (i.e., fast subthreshold frequency and lower spiking frequency). By virtue of their coherent inhibition, sparsely synchronized suprathreshold neurons suppress noisy activities of subthreshold neurons. Hence, only the coherent fast subthreshold hopping phase appears in the individual potentials of subthreshold neurons. The inhibitory coherence has  been characterized in terms of the ``statistical-mechanical'' coherence measures based on spikes and correlations. The degree of inhibitory coherence  was thus found to increase as $P_{supra}$ is increased for $P_{supra}>P_{supra}^*$. The effect of sparse random synaptic connectivity on the inhibitory coherence has also been  investigated, and emergence of inhibitory coherence has thus been found to persist only if $M_{syn}$ is larger than a threshold value $M_{syn}^*$. Finally, the universality of heterogeneity-induced inhibitory coherence has been confirmed in a population of canonical type-I QIF neurons. This kind of heterogeneity-induced weak inhibitory coherence which emerges from sparsely synchronized oscillations of suprathreshold neurons might be associated with cortical rhythms with irregular and sparse neural firings which contribute to cognitive functions such as information integration, working memory, and selective attention. Impaired sparse synchronization is believed to be related to mental disorders (e.g., schizophrenia and autism) \cite{W_Review}.

\begin{acknowledgments}
S.Y.K. completed the final version of the manuscript during his visit to the Jeju National University, and he thanks Profs. D. Kang and D. Kim for their hospitality. This research was supported by the Basic Science Research Program through the National Research Foundation of Korea funded by the Ministry of
Education, Science and Technology (2010-0015730).
\end{acknowledgments}

\end{document}